\documentclass[twocolumn,showpacs,amsmath,amssymb]{revtex4}

\usepackage{graphicx}
\usepackage{dcolumn}
\usepackage{bm}

\newcommand{\ket}[1]{|#1\rangle}

\begin{document}

\title{Dynamical Localization in Quasi-Periodic Driven Systems}

\author{G. Abal}
\email{abal@fing.edu.uy}

\author{R. Donangelo}
 \altaffiliation[Permanent address: ]{Instituto de F\'{\i}sica,\\ 
Universidade Federal do Rio de Janeiro,\\
C.P. 68528, 21945-970 Rio de Janeiro, Brazil}

\author{A. Romanelli}

\author{A.C. Sicardi Schifino}
 \altaffiliation[Also at ]{Instituto de F\'{\i}sica, Facultad de Ciencias.}

\author{R. Siri}

\affiliation{Instituto de F\'{\i}sica, Facultad de Ingenier\'{\i}a, 
Universidad de la Rep\'ublica\\
CC 30, CP 11000, Montevideo, Uruguay}

\date{\today}

\begin{abstract} 
We investigate how the time dependence of the Hamiltonian determines the
occurrence of Dynamical Localization (DL) in driven quantum systems with two
incommensurate frequencies. If both frequencies are associated to  impulsive
terms, DL is permanently destroyed. In this case, we show that the evolution 
is similar to a decoherent case. On the other hand, if both frequencies are
associated to  smooth driving functions, DL persists although on a time scale
longer than in the periodic case. When the driving function consists of a
series of pulses of duration $\sigma$, we show that the localization time
increases as $\sigma^{-2}$ as the  impulsive limit, $\sigma\rightarrow 0$, is
approached. In the intermediate case, in which  only one of the frequencies is
associated to an impulsive term in the Hamiltonian, a transition from a
localized to a delocalized dynamics takes place at a certain critical value of
the strength parameter. We provide an estimate for this critical value, based
on analytical considerations. We show how, in all cases, the  frequency
spectrum of the dynamical response can be used to understand the global
features  of the motion. All results are numerically checked.
\end{abstract}

\pacs{PACS: 05.45.+b, 03.65.-w, 24.60.Lz, 72.15.Rn}

\maketitle

\section{Introduction}
\label{sec:intro}

Dynamical Localization (DL), discovered numerically in 1979 \cite{CCI79}, has
attracted a great deal of attention after its experimental realization
in samples of cold atoms interacting with a far--detuned standing wave of laser
light \cite{exp_stand_wave}. When the light field is switched on and off
periodically, the system can be modeled by the Kicked Rotor Hamiltonian in a
regime in which quantum effects are important \cite{exp_qkr, Ammann}. This
notable series of experiments confirmed previous expectations regarding DL in
periodically driven quantum systems. More recently, it has been reported
\cite{Delande} that the addition of a second driving frequency, incommensurate
with the first, results in the destruction of DL in the experimentally
accessible time scale. The important conceptual issue of whether the addition
of a second incommensurate frequency permanently destroys DL or just causes a
substantial increase in the localization time, cannot be resolved
experimentally. On the other hand, numerical experiments alone are
intrinsically unable to distinguish between a very large increase in
localization time and an effective suppression of DL. Thus, some theoretical
insight on DL for non-periodically driven systems is presently required in
order to provide an answer to this kind of questions. 

Most theoretical work on the subject has dealt with the special case of
periodically driven systems \cite{Haake, Izrailev}. In comparison, little is
known on the quantum dynamics of driven systems when the external field is not
periodic. In this work we present a basic theoretical framework which may lead
to a deeper understanding of DL and quantum diffusion in quasiperiodic systems.
Our approach focuses in the relation between the density of the Fourier
spectrum of the dynamical response and the localized or diffusive character of
the quantum motion. As we shall see, the importance of the impulsive or
non-impulsive character of the driving has been underestimated in the past,
since it is at least as important for DL as the number of independent
frequencies in the Hamiltonian. Our conclusions are supported by numerical
simulations of several characteristic systems. 

In Section~\ref{sec:qkr2f}, a Kicked Rotor with two driving frequencies is
introduced and it is shown that DL is impossible in this system unless the
frequencies are commensurable. We also introduce here an energy balance which
will be of great importance in the following sections.  In
Section~\ref{sec:nonimp}, a quasiperiodic driven rotor without impulsive terms
in the Hamiltonian is discussed. It is shown that, in this kind of systems, DL
takes place for arbitrary driving strengths provided that the classical analog
is chaotic. In Section~\ref{sec:mod}, an intermediate system, the modulated
Kicked Rotor, is considered in detail. In this kind of multi-frequency systems,
one driving frequency is associated with the impulsive term and the other with
a smoothly varying modulation factor. This system has a transition, from
localized to diffusive dynamics, for kicking strengths above a certain
threshold. We show how this transition can be understood within our theoretical
framework and provide a concrete estimate for the threshold value for the
particular example that we consider. Finally, in Section~\ref{sec:conclusion}
we present a unifying discussion of these results and summarize our
conclusions.

\section{Two-frequency Quantum Kicked Rotor}
\label{sec:qkr2f}

Let us consider a Quantum Kicked Rotor (QKR) to which a second series of
periodic delta-kicks is applied. External kicks occur at times $t=nT_1$ and
$t=mT_2$ respectively ($n,m$ integers) and we write the Hamiltonian as   
\begin{equation} 
H=\frac{P^{2}}{2I}+\cos\theta\left[
K_{1}\sum _{n=1}^{\infty } \delta (t-nT_{1})+K_{2}\sum _{m=1}^{\infty }
\delta (t-mT_{2})\right].
\label{qkr2f_ham}     
\end{equation} 
where $I$ is the moment of inertia of the rotor, $P$ the angular
momentum operator and $K_1$ $(K_2)$ the strength parameter for the series of
kicks of periods $T_1$ $(T_2)$. The rational or irrational character of the
ratio $r=T_2/T_1$ determines whether the Hamiltonian (\ref{qkr2f_ham}) is
periodic or quasi-periodic.  

In the angular momentum representation, $P\ket{\ell}~=~\ell\hbar\ket{\ell}$,
the wavevector is 
$\ket{\Psi(t)}~=~\sum_{\ell=-\infty}^{\infty}a_{\ell}(t)\ket{\ell}$  and the
average energy is   $E(t)=\left\langle\Psi\right| H\left|\Psi \right\rangle=
\sum_{\ell=-\infty }^{\infty }E_{\ell}\left| a_\ell (t)\right|^{2}$,  where
$E_{\ell}=\ell^2\hbar^2/2I$ are the eigenvalues of $P^2/2I$.  As in the case of
the QKR, a quantum map 
\begin{equation}
\label{mapa}
a_{\ell}(t_{n+1})=\sum _{j=-\infty }^{\infty }i^{-(j-\ell)}
e^{-iE_{j}\Delta t_{n}/\hbar }\, J_{j-\ell}(\kappa _{n})a_{j}(t_{n})
\end{equation}
is readily obtained from the Hamiltonian (\ref{qkr2f_ham}).
In Eq.~(\ref{mapa}), we refer to the instant immediately after the $n^{th} $ kick
as $t_{n}$ and to the time interval between two consecutive kicks as $\Delta
t_{n}\equiv t_{n+1}-t_{n}$. The argument of the $k^{th}$ order cylindrical
Bessel function, $J_{k}$, is the dimensionless kick strength $\kappa_n\equiv
K_n/\hbar$ which, for this system, takes only the two values ($K_1/\hbar$ or
$K_2/\hbar$) depending on the kind of the $(n+1)^{th}$ kick. We will use $T\equiv T_1$ 
as the unit of time. 

After a straightforward calculation involving the map (\ref{mapa}), the energy
increase due to the $ (n+1)^{th} $ kick can be expressed as  
\begin{equation}
\label{ene1}
E(n+1)-E(n)= \frac{\hbar ^{2}}{2I}\left[\frac{\kappa_n^{2}}{2}+\Gamma_n\right]
\end{equation}
with $\Gamma_n\equiv$
\begin{eqnarray} 
2\kappa_n{\mathcal{I}}m \sum _{j=-\infty }^{\infty }(j+\frac 12)\,
a_{j}(t_{n})a^{*}_{j+1}(t_{n}) \, 
e^{-\frac{i}{\hbar}(E_{j+1}-E_j)\Delta t_{n}}&&\nonumber\\
-\frac{\kappa_n^{2}}{2}{\mathcal{R}}e
\sum_{j=-\infty }^{\infty } a_{j}(t_{n})a^{*}_{j+2}(t_{n})\,
e^{-\frac{i}{\hbar}(E_{j+2}-E_j)\Delta t_{n}}.&& \label{ene1a}
\end{eqnarray} 

The ensemble average of Eq.~(\ref{ene1}) is proportional to the diffusion
rate in angular momentum, $D_n~=~\overline{\Delta P_n^2}~/~T$, if the kick
number $n$ is used as a measure of time. The amplitude--independent term in 
Eq.~(\ref{ene1}) corresponds to the  quasi-linear approximation to the
classical diffusion coefficient \cite{Ott},  
\begin{equation}
D_{ql}=\frac{\overline{\kappa^2}}{2}, \label{Dql} 
\end{equation} 
where $\overline{\kappa^2}$ stands for the average value of $\kappa_n^{2}$.
The remaining terms, grouped as $\Gamma_n$ in Eq.~(\ref{ene1}), depend on
the wavevector $\{a_\ell\}$ and on the time interval between kicks. In sum,
the energy balance in Eq.~(\ref{ene1}), contains two qualitatively different
terms: one due to classical diffusion and the other ($\Gamma_n$) associated
to quantum interference effects. 

If the terms in $\Gamma_n$ have random phases, the sums are decoherent, and
thus have negligible mean values. Therefore, $\overline{\Gamma_n}=0$ and
the system mimics a classical evolution, and the average energy increases
linearly with the number of kicks with a slope given by (\ref{Dql}). On the
other hand, when DL takes place, the average value of the sums in $\Gamma_n$
must cancel out the independent term in Eq.~(\ref{ene1}) , so 
$\overline{\Gamma_n}=-\overline{\kappa^2}/2$. This can happen only if these
sums are coherent. Thus, the long time persistence of correlations among the
amplitudes $a_\ell$, taken at a given time, is a necessary condition
for DL to take place. We emphasize this rather obvious fact because it plays a
fundamental role in the discussions in the rest of this paper. 

\begin{figure} 
\includegraphics[scale=0.6, angle=-90]{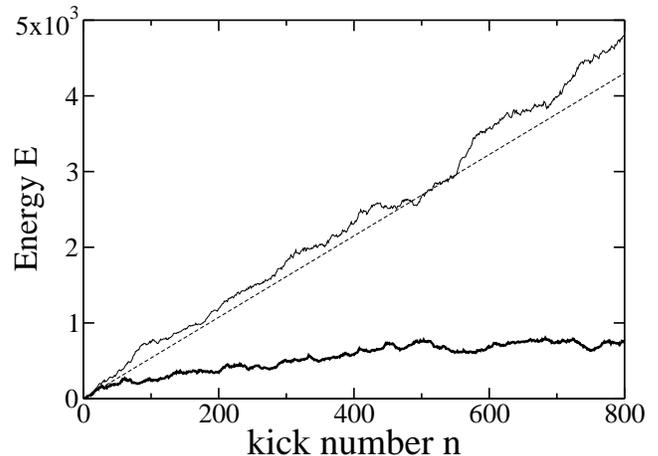} 
\caption{\label{fig:qkr2f}
\footnotesize Energy (in units of $\hbar^2/2I$) as a function of the
kick number, $n$, for the two--frequency quantum Kicked Rotor. The evolution
was generated from the map~(\ref{mapa}) with $\kappa = 3.279$ and $\xi =
1.525$. Two values of the period ratio $r\equiv T_{2}/T_{1}$ are
shown: $r=3/2$ (thick line) and $r=\sqrt{2}$  (thin line). The dashed line
corresponds to the quasi-linear approximation, Eq.~(\ref{Dql}). 
The initial state was taken to be $\ket{\ell =0}$.} 
\end{figure} 

We have performed a numerical study of the two-frequency quantum Kicked Rotor,
described by the map~(\ref{mapa}) with $K_1 = K_2$. We write the two
dimensionless parameters of the standard QKR model as  $\kappa = K/\hbar$ and
$\xi = \hbar T/I$. The period ratio, $r\equiv T_2/T_1$, is an additional
parameter present in the  two-frequency version. 

\begin{figure}[t] 
\includegraphics[scale=0.6, angle=-90]{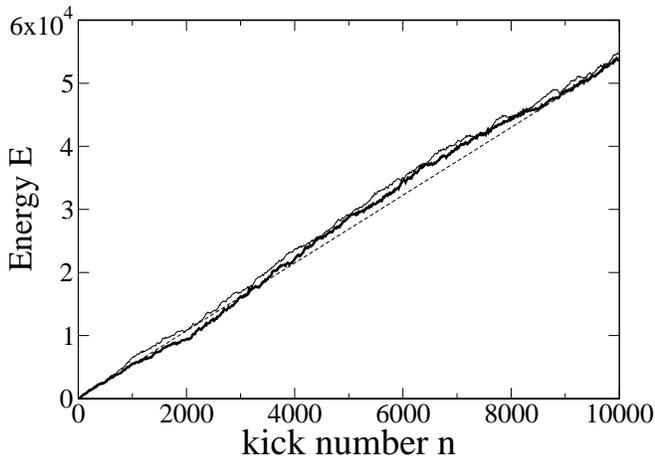} 
\caption{\label{fig:random}
\footnotesize 
Energy (in units of $\hbar^2/2I$) as a function of the kick number, 
$n$, for the Random Rotor defined in the text (thick line) and the
two-frequency Kicked Rotor with $r=\sqrt{2}$ (thin line). 
The dashed line, parameters, and initial conditions are the same as those in 
Figure~{\ref{fig:qkr2f}}.} 
\end{figure}

In Figure~\ref{fig:qkr2f}, we show the evolution of the energy of the rotor for
a rational and irrational ratio of the two periods $r\equiv~T_2/T_1$. 
In the first case, the behavior is similar to the usual one-frequency Kicked Rotor
and the energy initially increases in a diffusive way and then, after a characteristic
time, it localizes.
In the second case, the results strongly suggest that the diffusive behavior continues
indefinitely and that in this case dynamical localization does not take place.
In fact, we have checked that the energy increases at a rate consistent with
the quasi-linear approximation to the classical diffusion coefficient,
Eq.~(\ref{Dql}), for at least $10^4$ kicks (see Figure~\ref{fig:random}).

The results obtained for rational $r$ could be expected because, in this case,
the system is equivalent to a Kicked Rotor with a periodic train of pulses. For
an irrational $r$, our results can be better understood  by considering a
Random Rotor, defined as a Kicked Rotor for which kicks of fixed strength
$\kappa$ are applied at uniformly distributed random time intervals. In this
case, the map (\ref{mapa}) still holds, but the time interval between
consecutive kicks $\Delta t_{n}$, is now a random variable uniformly
distributed in $[0,T]$. In Figure~\ref{fig:random}, we compare the long time
evolution of the average energy of this  Random Rotor with the corresponding
quasiperiodic Kicked Rotor.  The evolution of the energy is essentially the
same in both cases and this suggests that the underlaying dynamics is very
similar. 

In the case of the quasiperiodic QKR, the time sequence $\Delta t_{n}$ is
obtained from a systematic rule, once the period ratio $r$ has been specified.
In spite of this, all time intervals in this sequence occur only once and as
$n\rightarrow\infty$ the values of $\Delta t_{n}$ are dense and uniformly
distributed in $[0,T]$, as is the case for random time intervals. In both
cases, numerical results confirm that $\overline{\Gamma_n}\approx 0$ and that 
classical-like diffusion takes place indefinitely. The effect of these time
intervals on the dynamics can be made more explicit by rewriting the map
(\ref{mapa}) in terms of the initial condition,  

\begin{widetext}
\begin{equation}
\label{telescopico} 
a_{\ell}(t_{n+1})=\sum _{j_{1},j_{2}\ldots j_{n+1}}i^{-(j_{1}-\ell)} 
e^{ -\frac{i}{\hbar }\sum _{m=1}^{n+1}E_{j_m}\Delta t_m} \,  
J_{j_{n+1}-\ell}(\kappa_n )J_{j_n -j_{n+1}}(\kappa_{n-1} )\ldots
J_{j_{1}-j_{2}}(\kappa_0)\, a_{j_{1}}(t_0). 
\end{equation} 
\end{widetext}

This expression can be applied to any of the systems discussed so far
(periodic, quasiperiodic and random). The only difference between the localized
and the  diffusive cases is in the sequence of time intervals, $\Delta t_{n}$.
In the familiar periodic case, when all time intervals are the same, it is well
known that the amplitudes $a_\ell$ are exponentially localized.  When they
are substituted in Eq.~(\ref{ene1}) a coherent sum results in $\Gamma_n$,
which, as  mentioned before, cancels on the average the classical diffusion
coefficient and results in a null mean energy increase. On the other hand, in
the quasiperiodic case, the phases
$-\frac{i}{\hbar}\sum_{m=1}^{n+1}E_{j_m}\Delta t_m$ (mod $2\pi$), appearing 
in Eq.~(\ref{telescopico}), form a dense, pseudorandom set in $[0,2\pi]$. When
the resulting amplitudes, $a_\ell$, are  substituted in Eq.~(\ref{ene1}),
they result in a negligible decoherent sum in $\Gamma_n$ and the classical 
diffusion coefficient is obtained.

It is interesting to remark that the previous discussion could have been
expressed equally well in terms of the Fourier frequencies associated to the
dynamics. In the quasiperiodic case, the time intervals between consecutive
kicks are of the form  
\begin{equation} 
\label{time_int} 
\Delta t_n=|pT_1 - qT_2|=\frac{2\pi}{\omega_1\omega_2}|p\omega_2-q\omega_1| 
\end{equation} 
where $p$ and $q$ are arbitrary integers and $\omega_{1,2}~=~2\pi/T_{1,2}$. 
These time intervals are dense in $[0,T]$. The corresponding frequencies, 
$\omega_{pq}~\equiv~|p\omega_2~-~q\omega_1|$, also form a dense set in
$[0,\omega_2]$. In other words, all frequencies are relevant for the dynamics 
and the time evolution of the energy mimics the classical chaotic diffusion, 
because the average separation between adjacent frequencies, $\Delta\omega$ is
null. It is well known that in the periodic QKR, DL is related to the discrete
nature of the quasienergy spectrum or, equivalently, to the frequency spectrum
of the dynamical response.  In fact, this spectrum is discrete is spite that a
classical chaotic system has a continuous frequency spectrum. However, due to
the Uncertainty Principle, this discreteness does not manifest itself until a
finite time of the  order of $1/\Delta\omega$, where $\Delta\omega$ is the
average separation between adjacent frequencies. For shorter times, the
dynamical evolution ``mimics'' classical diffusion, {\it i.e.} quantum
diffusion takes place. At larger times, for which the discrete nature of the
spectrum becomes manifest, the motion is exponentially localized
\cite{Izrailev}. A similar argument for the quasiperiodic two--frequency Kicked Rotor leads us to conclude that in this case, in which $\Delta\omega =0$, the localization time is infinite or, alternatively, that the addition of a second incommensurable frequency permanently destroys DL in impulsive systems, such as the QKR. 

\section{Non-impulsive systems }
\label{sec:nonimp}

In the previous section, we have shown that quantum diffusion takes place for 
ever in a quasiperiodic Kicked Rotor in which the driving function has two 
impulsive components. In this section, we consider the diffusive properties of 
smoothly driven quantum systems with two frequencies $\omega_{1}$ and
$\omega_{2}$.  

As an example, consider a rotor in which the driving force consists of two
series of periodic narrow pulses. Such a system is described by the
Hamiltonian
\begin{equation}  
\label{qfr2f_ham}  H=\frac{P^{2}}{2I}+\cos \theta \,
\left[K_{1}f_{1}(t)+K_{2}f_{2}(t)\right],   
\end{equation}  
where $f_{1}(t)=f_{1}(t+T_{1}) $ and $ f_{2}(t)=f_{2}(t+T_{2}) $ are smooth
periodic functions of time. If $ r=T_{2}/T_{1} $ is rational, the Hamiltonian
(\ref{qfr2f_ham}) is periodic and DL is expected to occur.  If $r$ is an
irrational number, recent experimental results \cite{Delande} show that DL is
destroyed or at least the localization time is increased by an order of
magnitude. In view of the discussion of the previous section, one might expect
that unlimited diffusion would result also in this case. However, as we show
below, DL  persists in the quasiperiodic, non-impulsive case.

In order to fix ideas, we specify each of the driving functions as a periodic
sequence of Gaussian pulses of characteristic width $\sigma$, so that for 
$s=1,2$
\begin{equation}
\label{fgauss}  
f_{s}(t)=\frac{1}{\sigma \sqrt{{2\pi }}}
\sum^{\infty}_{n=-\infty}
e^{-\frac{\left(t-nT_s-\Phi_s\right)^2}{2\sigma^2}}. 
\end{equation}  
Without loss of generality we choose $\Phi_1 =0$ and $\Phi_2 =\Phi$. 
Thus, the driving function in (\ref{qfr2f_ham}) consists of a superposition 
of pulses of strength $K_{s}$ which occur at times $ T_{1},2T_{1},\ldots$ and 
$\Phi+T_{2},\Phi+2T_{2},\ldots$ In the limit $\sigma\rightarrow 0$, the pulses 
reduce to delta functions and the Hamiltonian (\ref{qfr2f_ham}) reduces to 
(\ref{qkr2f_ham}), describing a two--frequency kicked rotor. As discussed in 
Section~\ref{sec:qkr2f}, if $r$ is irrational, this system shows diffusion for
ever. 

At this point, there is one important difference to bear in mind: in the
classical Kicked Rotor, it is well known that for large $K$ (in practice, 
$K\agt 5$ suffices) all KAM surfaces are destroyed and the energy increase 
is unbounded \cite{Ott,L+L}.  In contrast, for finite $\sigma $, the momentum 
spread is limited by the existence of KAM surfaces for all values of $K$. 
However, this upper bound in momentum space increases with K in a predictable 
form \cite{QFR}. We have checked that the classical phase space accessible in 
the time scales considered here is completely chaotic, so
that these KAM boundaries do not affect our results.

\begin{figure}
\includegraphics[scale=0.6, angle=-90]{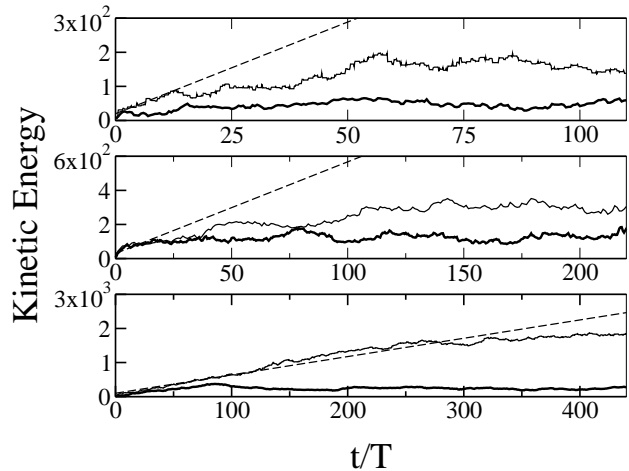}
\caption{\footnotesize
Kinetic energy (units of $\hbar^2/2I$) of the quantum Forced Rotor 
described by Eq.~(\ref{qfr2f_ham}) with gaussian driving functions given by
Eq.~(\ref{fgauss}). Three characteristic pulse widths are shown: upper panel 
$ \sigma/T =0.05$, medium panel $ \sigma /T=0.03$ and lower panel $ \sigma /T
=0.01$. The other parameters are fixed at $\kappa_1~=~\kappa_2~=~3.279$ and
$\xi~=~1.525$. The thick lines correspond to the periodic case with $r=3/2$,
the thin lines to the quasiperiodic case with $ r=\sqrt{{2}}$ and the dashed
lines to the energy increase predicted by the quasilinear approximation to the
classical diffusion coefficient, Eq.~(\ref{Dql}). In all cases, the initial
state was a Gaussian packet in the momentum representation, centered at 
$\ell = 0$.}
\label{fig:qfr2f}
\end{figure}

The Schr\"odinger equation for the Hamiltonian (\ref{qfr2f_ham}) can be
expressed in the angular momentum representation as
\begin{equation}
\label{Schro_qfr}
\dot{a}_{n}+\frac{iE_{n}}{\hbar }a_{n}+\frac{i}{2}\left[ \kappa _{1}f_{1}(t)+
\kappa _{2}f_{2}(t)\right] \left( a_{n-1}+a_{n+1}\right) =0.
\end{equation}
We have numerically integrated this equation and calculated the average energy
as a function of time, for a small (but finite) pulse width $\sigma$. Our
results are shown in Figure~\ref{fig:qfr2f}, for several values of $\sigma$. As
expected, they depend strongly on the rational or irrational character of the
ratio $r=T_{2}/T_{1}$.  In the case of two commensurable frequencies (periodic
driving) the rotor localizes after a few kicks. When the two frequencies are
incommensurate, the system also localizes but in a much longer time scale. The
persistence of DL found here is in striking contrast with the case of impulsive
driving discussed in Section~\ref{sec:qkr2f}. The localization time for
quasiperiodic driving increases as $\sigma\rightarrow 0$, approximately as
$\sigma^{-2}$.  We explain this dependence below, but here we note that the 
localization time becomes infinite in the impulsive limit.

We can understand these results by considering the Fourier transform of the
driving function $f(t)~=~\kappa_1f_1(t)~+~\kappa_2f_2(t)$, given by 
\begin{eqnarray}
F(\omega)&=& \frac{e^{-\omega^2\sigma^2/2}}{\sigma\sqrt{2\pi}} 
\left[ \kappa_1
\sum_{m=-\infty}^\infty \delta(\omega - m\omega_1)\right.\nonumber\\
&&\qquad +\left.\kappa_2\sum_{n=-\infty}^\infty \delta(\omega - n\omega_2)\right].
\label{Fourier_transf} 
\end{eqnarray} 

The dynamical response which emerges from Eq.~(\ref{Schro_qfr}) involves the 
differences of the frequencies in the Fourier spectrum of the driving function
$f(t)$, that is, $\omega_{nm}~=~|n\omega_2-m\omega_1|$. We have previously
introduced these frequencies in  Eq.~(\ref{time_int}).  As $\sigma\rightarrow
0$ and the Gaussian modulation factor in (\ref{Fourier_transf}) becomes unity,
all harmonics of $\omega_1$ and $\omega_2$ are equally important in
(\ref{Fourier_transf}) and all the differences $\omega_{nm}$ appear with
equal weights in the dynamical response. In this case, the frequency spectrum
of the response is dense and there is no DL (see Figure~\ref{fig:qkr2f}). For
finite $\sigma$, the Gaussian modulation factor effectively suppresses all
high harmonics $(\omega > 1/\sigma)$ from the Fourier spectrum of the driving
function. Then, only a finite number of linear combinations
$\omega_{nm}$ are important in the dynamical response, so the spectrum is
effectively discrete and DL takes place. 

The above mentioned $\sigma^{-2}$ dependence of the localization time can be
understood recalling that, according to the argument presented at the end of
Section~\ref{sec:qkr2f}, this time is inversely proportional to the average
separation between adjacent frequencies, $\Delta\omega$. This quantity is
inversely proportional to the number of significant frequencies, $\omega_{nm}$,
that appear in the dynamics of the system. Then, the localization time is  
proportional to this number of relevant frequencies. Since only $\sim 1/\sigma$
harmonics of each fundamental frequency enter in the dynamics, the number of
significant frequencies, $\omega_{nm}$, and the localization time, both
increase as $1/\sigma^2$ as $\sigma$ is reduced.   We note that a similar
argument has been used in the context of periodically  driven systems to
establish the proportionality of the localization time to the number of
quasienergies which  participate in the dynamics \cite{Izrailev}.

Another example of a non--impulsive, driven quantum system is provided by a particle confined in an infinite square well with a periodically changing width,
$L(t)$. This system, known as the Fermi Accelerator, shows DL when its classical counterpart is chaotic \cite{Abal1, Abal2}. We have considered a 
quasiperiodic version of this system in the context of Nuclear Dissipation
Theory and found that the localization time increases by an order of magnitude but DL still takes place \cite{NPA1,NPA2}. An analysis completely analogous to the one presented above explains the persistence of DL in the quasiperiodic Fermi Accelerator as a consequence of the discreteness of the effective frequency spectrum of the dynamical response. 

\section{Modulated Kicked Rotor}
\label{sec:mod}

In Section~\ref{sec:qkr2f}, we have considered a Kicked Rotor driven by two
impulsive driving functions of different frequencies and showed that DL does
not take place when the frequency ratio is irrational. In contrast, in
Section~\ref{sec:nonimp}, we presented two examples of smoothly driven
quasiperiodic systems in which DL persists, although in a longer time scale
than in the periodic case. We now consider an intermediate case in which the
driving function has two frequencies but only one of them has an impulsive
character, while the other one is associated with a smooth function of time
that multiplies the impulsive term. 

\begin{figure}
\centerline{\includegraphics[scale=0.55,angle=-90]{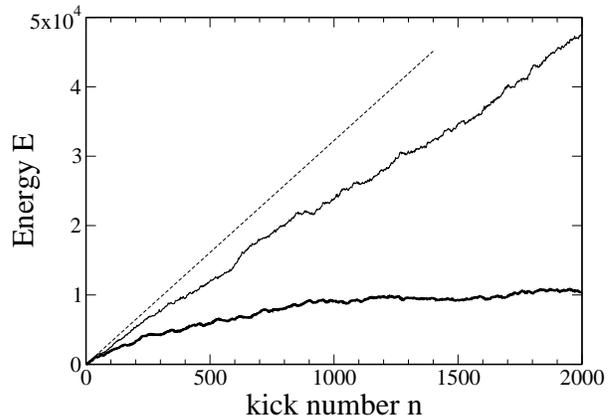} }
\caption{\footnotesize
Energy (in units of $\hbar^{2}/2I$) of the modulated Quantum Kicked Rotor,
Eq.~(\ref{mod_ham}),  as obtained from the quantum map ~(\ref{mapa}) for
$\kappa-n$ given by  Eq.~(\ref{kappa_{n}}).  The thick line corresponds to the
periodic case with $r=3/2$ and the thin line to the quasiperiodic case with
$r=\sqrt{{2}}$. The initial state is the ground state of the unperturbed
system, $\ket{\ell = 0}$. The dashed line has a slope given by the quasilinear
approximation to the classical diffusion coefficient, Eq.~(\ref{Dql}), with the
average kicked strength from Eq.~(\ref{kappa_eff}) .  The other parameters are
$\kappa = 13.114$ and $\xi=1.525$.}
\label{fig:mod_2f}
\end{figure}

Two examples of such systems have been discussed in \cite{CGS89,
Shepelyansky}.  Here, we consider a simple periodic modulation so that the
rotor is described by the Hamiltonian 
\begin{equation}
\label{mod_ham} 
H=\frac{P^{2}}{2I}+K\cos \theta \cos ^{2}(2\pi t/T_{2})
\sum_{n=1}^{\infty }\delta (t-nT_{1}). 
\end{equation} 
which corresponds to a Kicked Rotor with a fixed interval $T\equiv T_1$ between kicks 
and a kick strength modulated by a function of period $T_2$. We refer to the
quantum version of this system as the modulated Quantum Kicked Rotor.
The map~(\ref{mapa}) is still valid in this case if the time dependent kick strength is 
redefined as 
\begin{equation} 
\label{kappa_{n}} 
\kappa_{n}\equiv\kappa \cos ^{2}(2\pi n/r).
\end{equation} 

The evolution of the energy, obtained by iterating the map~(\ref{mapa}) for the
first $2000$ kicks, is shown in Figure~\ref{fig:mod_2f}. For rational $r$, the
energy localizes as expected, since in this case the system is periodic and
equivalent to a QKR. Localization is broken in the quasiperiodic case 
(irrational $r$) and, in what follows, we will focus our attention in this 
case. The dashed line
in Figure~\ref{fig:mod_2f} corresponds to the quasilinear approximation to the
diffusion coefficient, Eq.~(\ref{Dql}) calculated for the average
value of the squared kick strength,
\begin{equation} 
\label{kappa_eff} 
\overline{\kappa^{2}}\equiv\frac{1}{N}\sum^{N}_{n=1}\kappa_{n}^2=
\frac{3}{8}\kappa^{2}. 
\end{equation}
The quantum diffusion, shown in Figure~\ref{fig:mod_2f}, takes place at a slower
rate that the classical one. The reason for this slower quantum diffusion
rate will soon become apparent.

\begin{figure}
\centerline{\includegraphics[scale=0.6, angle=-90]{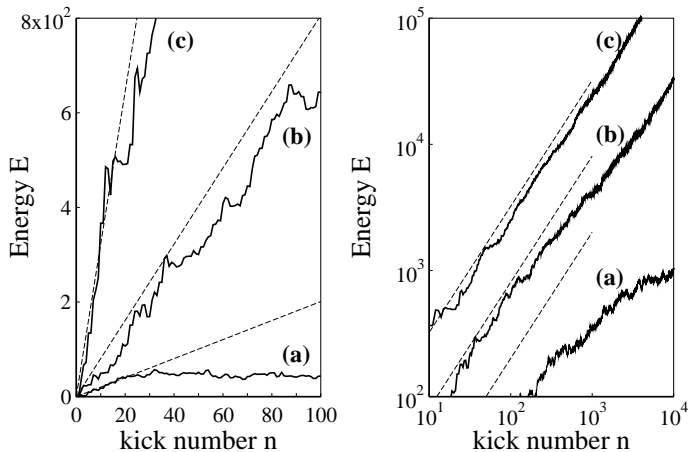}} 
\caption{\footnotesize
Energy (units of $\hbar^2/2I$) for the modulated Quantum Kicked Rotor as a
function of kick number, $n$, for three different kick strengths: (a) $\kappa
=3.28$, (b) $\kappa =6.56$ and (c) $\kappa =13.11$. The scale parameter has
been fixed at $\xi =1.525$. In all cases, the dashed line corresponds to the
quasilinear approximation to classical diffusion. In the right panel, the long
time behavior (note the log-log scales) is given. In the left panel, the
detailed short-time evolution corresponding to the boxed region in the right
panel is shown in a linear scale.}
\label{fig:mod_trans}
\end{figure}

As shown in the left panel of Figure~\ref{fig:mod_trans}, for small values of
$\kappa$ the dynamics is localized. As $\kappa$ is increased, the evolution of
the energy shows a transition between a localized and delocalized dynamics.
There is some critical value such that for $\kappa<\kappa_{crit}$, DL takes
place after a characteristic time, but for $\kappa >\kappa_{crit}$ quantum
diffusion persists for very long times. We have checked that the diffusion
continues for at least $10^4$ periods. Furthermore, as $\kappa\gg\kappa_{crit}$
the diffusion rate approximates the classical one. In fact, the same
diffusion rate is obtained for a series of kicks of random strengths,
obtained by replacing the ratio $t/T_2$ in Eq.~(\ref{mod_ham}) by a uniform
random variable in $[0,1]$. This kind of transition, between a localized and a
delocalized regime, has been reported in connection with other versions of the
modulated Kicked Rotor \cite{Shepelyansky,CGS89}. 

The existence of this transition, as well as the fact that the quantum
diffusion rate is lower than the classical one, can both be understood from a
detailed inspection of the map (\ref{mapa}). We start by noting that since in
this case $\Delta t_n=T$, the only time dependence in the coefficients in the
rhs of (\ref{mapa}) appears in the argument of the Bessel function. The 
frequencies introduced in the dynamics by this time dependence can be made
explicit by recalling the definition of the Bessel function
\begin{equation}
J_\nu(\kappa_n)=\sum_{k=0}^\infty (-1)^k \frac{1}{k!(\nu + k)!} 
\left(\frac{\kappa_n}{2}\right)^{2k+\nu}.
\label{bessel}
\end{equation}

If the complex form of Eq.~(\ref{kappa_{n}}) for $\kappa_n$ is substituted in
Eq.~(\ref{bessel}), a series of the form $\sum_p c_p e^{ip\omega t}$ is
obtained, in which the coefficients $c_p$ decay at a rate which depends on
$\kappa$. When this series expansion for $J_\nu$ is substituted in
Eq.~(\ref{telescopico}), after $n$ kicks each frequency will introduce $n$
harmonics in Eq.~(\ref{telescopico}). More generally, if there are $q$
relevant frequencies in Eq.~(\ref{bessel}), they introduce $q^n$ frequencies in
Eq.~(\ref{telescopico}). Furthermore, when the energy is calculated from
Eq.~(\ref{ene1}), these frequencies produce phases that interfere between
themselves giving rise to a localized or delocalized dynamics, depending on the
value of $\kappa$. An estimate of the critical value, $\kappa_{crit}$, can be
obtained by comparing the average over several kicks, $\overline{\kappa_n/2}$, 
appearing in Eq.~(\ref{bessel}), with unity, so that $\kappa_{crit} \approx 4$. 

This value is consistent with our numerical observations, as implied by
Figure~\ref{fig:mod_trans}. When $\kappa\gtrsim 4$, many frequencies are
present with non--negligible amplitudes in the Bessel function (\ref{bessel}).
These frequencies result in a dense response spectrum when they are
``amplified'' in Eq.~(\ref{telescopico}). In this case the sums in
Eq.~(\ref{ene1a}) are incoherent,  $\bar\Gamma_n\approx 0$ and quantum
diffusion takes place at a rate that gradually approaches the classical one as
$\kappa$ is increased. On the other hand, if $\kappa$ becomes smaller, the
amplitudes $c_p$ decay faster, fewer frequencies are relevant in
Eq.~(\ref{bessel}) and the response spectrum has a smaller density. This
produces quantum diffusion at a reduced rate as compared to the classical rate.
At some value $\kappa\lesssim 4$, there is a qualitative change in the dynamics
as the response spectrum undergoes a topological change from dense to discrete.
Then, the sums in Eq.~(\ref{ene1}) are coherent,  $\bar\Gamma_n = -\kappa^2/2$
and DL takes place.  

\section{Conclusions}
\label{sec:conclusion}

All the results presented in this work can be understood in terms of the
general argument presented in the last paragraph of Section~\ref{sec:qkr2f},
based on Heisemberg's Uncertainty Principle. According to it, a time of the
order of $1/\Delta\omega$ is required in order to resolve a separation
$\Delta\omega$ in the frequency domain. In particular, when the dynamical
response has a dense frequency spectrum, $\Delta\omega\rightarrow 0$, and the
quantum system mimics classical diffusion for arbitrarily long times. The
concrete mechanism resulting in the destruction of DL can be seen in the energy
balance,  Eq.~(\ref{ene1}), introduced in Section~\ref{sec:qkr2f}. In the case
of a dense frequency spectrum, the interference term of this equation is a
decoherent sum of null mean value. On the other hand, when the response
frequency spectrum has a discrete character, this sum is coherent and accounts
for DL, as in the QKR.

We have considered three different kinds of quasiperiodically driven systems in
Sections~II to IV. The differences in their dynamics can be understood in terms
of the dense or discrete character of the frequency spectrum of the dynamical
response. 

In the two-frequency Kicked Rotor, considered in Section~\ref{sec:qkr2f}, the
time intervals between kicks form a dense set and this produces a dense
frequency spectrum in the response. Thus, this system never localizes. In
smoothly driven quasiperiodic systems, such as the rotor driven by pulses of
duration $\sigma$, discussed in Section~\ref{sec:nonimp}, the effective
frequency spectrum has a discrete character. We have shown that
the average separation between frequencies, $\Delta\omega$, is in this case
proportional to $\sigma^2$. Then both, the number of relevant harmonics and the
localization time, increase as $1/\sigma^2$ as the impulsive limit
$\sigma\rightarrow 0$ is approached. In the intermediate case of a Modulated
Kicked Rotor, presented in Section~\ref{sec:mod}, the character of the response
spectrum depends on the kick strength parameter $\kappa$. As we have discussed, 
this parameter determines the number of significant linear combinations of the
fundamental frequencies that appear in the dynamical response. This explains
the existence of a threshold $(\kappa_{crit}\sim 4)$ below which DL takes
place. For $\kappa\gg\kappa_{crit}$, the diffusion rate approaches the
classical one.  

To conclude, we have established that quasiperiodically driven systems may
delocalize even in the absence of coupling with its environment. This is
possible when they are driven by impulsive terms with two or more
incommensurate frequencies. The additional incommensurate frequencies act as a
substitute for the coupling to a noisy environment. Furthermore, we have shown
that a strong causal connection exists between DL and the density of the
dynamical response spectrum. This spectrum can be used to characterize the
dynamics in an analogous form as the quasienergy spectrum in periodically
driven systems. Finally, we have shown that the impulsive or smooth character
of the driving terms of the Hamiltonian is as important for DL as the rational
or irrational character of the frequency ratio. 

Further work is required in order to understand how this considerations can be
extended to accommodate, for example, interactions with the environment.

\emph{We acknowledge the support of PEDECIBA and CONICYT-Clemente Estable
(project \#6026), RD acknowledges partial financial support from MCT/FINEP/CNPq
(PRONEX) under contract 41.96.0886.00.}


\begin{thebibliography}{99}

\bibitem{CCI79}G. Casati, B.V. Chirikov, F.M. Izrailev and J. Ford, Lect. Notes
Phys. \textbf{93}, 334 (1979). 

\bibitem{exp_stand_wave}F.L. Moore, J.C. Robinson, C. Bharucha, P.E. Williams
and M.G. Raizen, Phys. Rev. Lett. \textbf{73}, 2974 (1994) . 

\bibitem{exp_qkr}J.C. Robinson, C. Bharucha, F.L. Moore, R. Jahnke, G.A.
Georgakis, Q. Niu, M.G. Raizen and B. Sundaram, Phys. Rev. Lett. \textbf{74}, 3963 (1995);
J.C. Robinson, C.F. Bharucha, K.W. Madison, F.L. Moore, B. Sundaram, S.R.
Wilkinson and M.G. Raizen, Phys. Rev. Lett. \textbf{76}, 3304 (1996). 

\bibitem{Ammann} H. Ammann, R. Gray, I. Shvarchuck and N. Christensen, Phys.
Rev. Lett. \textbf{80}, 4111 (1998).

\bibitem{Delande}J. Ringot, P. Szriftgiser, J.C. Garreau and D. Delande,Phys.
Rev. Lett., \textbf{85}, 2741 (2000).

\bibitem{Haake} F. Haake, {\it Quantum Sugnatures of Chaos}, Springer Series in
Synergetics \textbf{54}, Springer--Verlag (1992).

\bibitem{Izrailev}F.M. Izrailev, Phys. Rep. \textbf{196,} 299 (1990). 

\bibitem{Ott}E. Ott, \textit{Chaos in Dynamical Systems}, Cambridge University
Press, (1993).

\bibitem{L+L}A. J. Lichtemberg and M.A. Lieberman, Regular and Stochastic
Motion, Springer-Verlag, New York (1983). 

\bibitem{QFR}G. Abal, R. Donangelo, A. Romanelli, A.C. Sicardi-Schifino, R.
Siri, in preparation (2001).

\bibitem{Abal1}G. Abal, A. Romanelli, A.C. Sicardi-Schifino, R. Siri and R.
Donangelo, Physica \textbf{A257}, 289 (1998). 

\bibitem{Abal2}G. Abal, A. Romanelli, A.C. Sicardi-Schifino, R. Siri and R.
Donangelo, Physica \textbf{A272}, 87 (1999). 87. 


\bibitem{NPA1}G. Abal, A. Romanelli, A.C. Sicardi Schifino, R. Siri and R.
Donangelo, Nucl. Phys. \textbf{A643} (1998) 30. 

\bibitem{NPA2}G. Abal, A. Romanelli, A.C. Sicardi Schifino, R. Siri and R.
Donangelo, Nucl. Phys. \textbf{A683} (2001) 279. 

\bibitem{CGS89}G. Casati, I. Guarneri and D.L. Shepelyansky, Phys. Rev. Lett.
\textbf{62} (1989) 345. 

\bibitem{Shepelyansky}D. Shepelyansky, Physica D \textbf{8}, 208 (1983). 
\end{thebibliography}
\end{document}